\documentclass[twocolumn,preprintnumbers,superscriptaddress,nofootinbib,aps,prd,floatfix]{revtex4}

\usepackage{datetime}
\usepackage{color}

\usepackage{footmisc,multirow}
\usepackage{subfigure}
\usepackage{amsmath,slashed}

\usepackage{graphicx}
\newcommand{\be}{\begin{eqnarray*}}
\newcommand{\ee}{\end{eqnarray*}}
\newcommand{\gl}[1]{(\ref{#1})}

\newcommand{\bee}{\begin{eqnarray}}
\newcommand{\eee}{\end{eqnarray}}
\newcommand{\beeq}{\begin{equation}}
\newcommand{\eeeq}{\end{equation}}
\newcommand{\cp}{${\cal{CP}}$}

\newcommand{\BR}{\text{BR}}

\renewcommand{\vec}{\bf}

\preprint{IPPP/14/37} \preprint{DCPT/14/74}

\begin{document}

\title{Limitations and Opportunities of Off-Shell Coupling
  Measurements}

\begin{abstract}
  Indirect constraints on the total Higgs width $\Gamma_h$ from
  correlating Higgs signal strengths with cross section measurements
  in the off-shell region for $p(g)p(g)\to 4\ell$ production have
  received considerable attention recently, and the CMS collaboration
  have published a first measurement. We revisit this analysis from a
  new physics and unitarity constraints perspective and conclude that
  limits on $\Gamma_h$ obtained in this fashion are not reliable
  unless we make model-specific assumptions, which cannot be justified
  at the current stage of the LHC programme. Relaxing the $\Gamma_h$
  interpretation, we discuss the merits of high invariant mass cross
  section measurements in the context of Higgs \cp~analyses, higher
  dimensional operator testing, and resolved new physics in the light
  of electroweak precision constraints beyond effective theory
  limitations. Furthermore, we show that a rather model-independent
  LHC constraint can be obtained from adapting the $gg\to 4\ell$
  analysis to the weak boson fusion channels at lower statistical
  yield.
\end{abstract}

\author{Christoph Englert} \email{christoph.englert@glasgow.ac.uk}
\affiliation{SUPA, School of Physics and Astronomy,\\ University of
  Glasgow, Glasgow G12 8QQ, United Kingdom\\[0.1cm]}

\author{Michael Spannowsky} \email{michael.spannowsky@durham.ac.uk}
\affiliation{Institute for Particle Physics Phenomenology, Department
  of Physics,\\Durham University, Durham DH1 3LE, United Kingdom\\[0.1cm]}

\maketitle

%%%%%%%%%%%%%%%%%%%%%%%%%%%%%%%%%%%%%%%%%%%%%%%%%%

\section{Introduction}
\label{sec:intro}
After the 2012 discovery~\cite{hatlas,hcms}, the ATLAS and CMS
collaborations have scrutinized the SM interpretation of the Higgs
candidate within the boundaries of the currently available data. A
strong resemblance of the particle's properties with the SM Higgs
expectation has emerged: it is likely to be a \cp~even scalar boson and
its ``signal strengths''
\begin{equation}
  \label{eq:signalstrength}
  \mu_{i,j}={\sigma_{h,i} \times \BR_j \over [\sigma_{h,i} \times \BR_j]^{\text{SM}}
  } \sim {\Gamma_i \Gamma_j 
    \over \Gamma_i^{\text{SM}} \Gamma_j^{\text{SM}}}{\Gamma_h^{\text{SM}}\over \Gamma_h}
\end{equation}
are in good agreement with the SM Higgs boson. $i,j$ in
Eq.~\gl{eq:signalstrength} refer to the different Higgs production and
decay modes that have been observed so far. For fully inclusive
measurements they can be related to the partial decay widths
$\{\Gamma_i\}$. ``Higgsistence'' has mainly been established from
gluon fusion, the largest Higgs production mechanism in the SM.

The apparent agreement of the measured quantities of
Eq.~\eqref{eq:signalstrength} with the SM predictions highlights the
question whether the discovered particle is indeed the Higgs boson as
predicted by the SM. 

On the one hand, unitarity largely constrains the bare couplings of
massive fermions and gauge bosons to \cp~even Higgs boson(s) in the
SM. If the absolute values of the Higgs candidate's couplings are
close to the SM predictions, there will be little room left for
resonant physics beyond the SM in {\it e.g.} the weak boson fusion
(WBF) channels, which is a direct probe of longitudinal gauge boson
scattering.

On the other hand, absolute values of couplings are difficult to infer
at hadron colliders since signal strength measurements involve
non-linear relations among the couplings and $\sigma \times \BR$
phenomenology leaves the total Higgs width as a flat direction in
coupling fits. This is usually overcome by making assumptions about
the total Higgs width in these fits~\cite{theo}, or, alternatively,
about the maximum coupling value of the Higgs candidate to gauge
bosons~\cite{spans} which is determined by the Higgs' gauge
representation. The biases that are introduced in either of these
approaches are far from being well-motivated at the current stage.

%%%%%%%%%%%%%%%%%%%%%%%%%%%%%%%%%%%%%%%%%%%%%
\begin{figure*}[!t]
  \begin{center}
    \hspace{-0.6cm}
    \parbox{0.5\textwidth}{
      \includegraphics[width=0.50\textwidth]{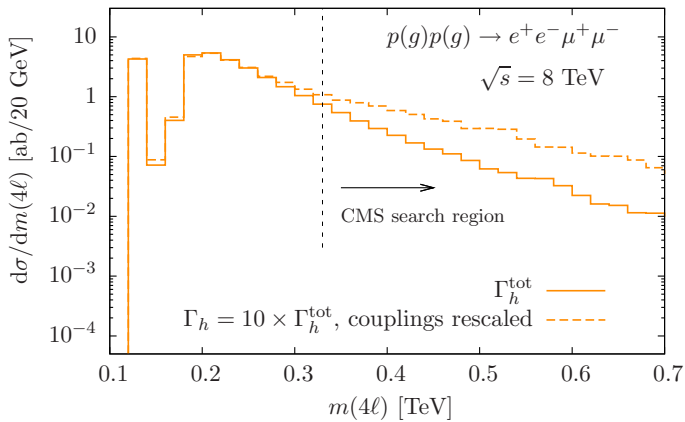}
    }
    \hspace{0.3cm}
    \parbox{0.4\textwidth}{ \vspace{1.5cm} \caption{Constraining the
        total Higgs width by fixing the signal strength (on-shell
        region) and measuring the cross section at large invariant
        $ZZ$ masses, keeping couplings in the on-shell and Higgs
        off-shell region fixed. Distributions are leading order, while
        keeping all quarks dynamical and the bottom and top quarks
        massive. We have chosen a minimal cut set $p_{T}(\ell)\geq
        10~\text{GeV}$, $|y(\ell)|\leq 2.5$, $\Delta R(\ell\ell')\geq
        0.4$.}
      \label{fig:zzsm}
      }
  \end{center}
\end{figure*}
%%%%%%%%%%%%%%%%%%%%%%%%%%%%%%%%%%%%%%%%%%%%%

Assuming $\Gamma_h\simeq \Gamma_h^{\text{SM}} \simeq 4~\text{MeV}$
skews coupling fits towards a parameter region that is oblivious of
the Higgs bosons' potential interplay with dark matter
phenomenology~\cite{abdel} and other phenomena that can be introduced
via well-motivated portal-type
interactions~\cite{portal,Englert:2011us}.

Assumptions about the Higgs $SU(2)_L$ representation are usually
limited to the $\bf{2}$'s due to the (accidental) custodial isopin
symmetry that preserves $T\simeq 1$ in (multi-) Higgs doublet
models. However, it is known that both current signal strength
measurements and electroweak precision constraints can be accounted
for in models with non-doublet Higgs fields~\cite{Englert:2013zpa} and
the complementing searches for Higgs exotics~\cite{Englert:2013wga}
necessary to rule out such an option are not available yet.

\bigskip

Obviously, a model-independent constraint on $\Gamma_h$ (or
$\BR(\text{invisible})$ if a particular model leaves production modes
unaltered) has a huge impact on BSM physics.\footnote{See {\it e.g.}
  Ref.~\cite{abdel} for a discussion of the invisible branching ratio
  measurements, {\it e.g.}~\cite{atlasinv}, in relation with dark
  matter phenomenology.} Hence, it is not surprising that the recent
proposal by Caola and Melnikov~\cite{melnikov} that interprets
off-shell cross section measurements of $pp\to
4\ell$~\cite{Kauer:2013qba} as a probe of $\Gamma_h$ has received
considerable
attention~\cite{ciaran,ciaran2,Coleppa:2014qja}.\footnote{Similar
  strategies~\cite{Dixon:2013haa} have been proposed for $h\to
  \gamma\gamma$~\cite{Dixon:2013haa2}.} Just recently CMS have
presented first results~\cite{cmswidth} using this strategy, claiming
$\Gamma_h < 4.2\times\Gamma_h^{\rm{SM}}$ at 95\% confidence level by
injecting a global Higgs signal strength $\mu\simeq 1$. The strategy
is sketched in Fig.~\ref{fig:zzsm}; and we give a quick outline to
make this work self-contained (for additional details
see~\cite{melnikov,cmswidth,ciaran2}):

%%%%%%%%%%%%%%%%%%%%%%%%%%%%%%%%%%%%%%%%%%%%%
\begin{figure}[!b]
  \begin{center}
    \includegraphics[width=0.47\textwidth]{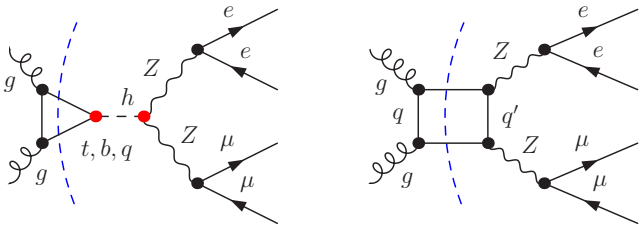}
    \caption{Representative Feynman diagram topologies contributing to
      $gg\to ZZ$ with leptonic $Z$ boson decays in the SM and theories
      with extended fermionic sectors.}
    \label{fig:feyzz}
  \end{center}
\end{figure}
%%%%%%%%%%%%%%%%%%%%%%%%%%%%%%%%%%%%%%%%%%%%%

As long as the narrow width approximation is applicable, the cross
section for the process $p(g)p(g) \to h \to Z Z^\ast \to 4\ell$ in the
the Higgs on-shell region scales as\footnote{We mainly focus on
  the final state $e^+ e^- \mu^+ \mu^-$ in the following. Generalizing
  our results to full leptonic $ZZ$ decays is straightforward due to
  negligible identical fermion interference.}
\begin{equation}
  \label{eq:onshell}
  \sigma_{h,g}\times \BR(h\to ZZ\to 4\ell) \sim {g^2_{ggh}\, g^2_{hZZ}
    \over \Gamma_h}\,,
\end{equation}
where we denote the relevant couplings by $g_X$. The dominant Feynman
diagram in this phase space region is the triangle of
Fig.~\ref{fig:feyzz}, the continuum contribution from $gg\to ZZ^\ast$
is highly suppressed and interference is
negligible~\cite{Kauer:2013qba}.

%%%%%%%%%%%%%%%%%%%%%%%%%%%%%%%%%%%%%%%%%%%%%
\begin{figure*}[!t]
  \begin{center}
    \hspace{-0.6cm}
    \parbox{0.50\textwidth}{
      \hspace{-0.6cm}\includegraphics[width=0.50\textwidth]{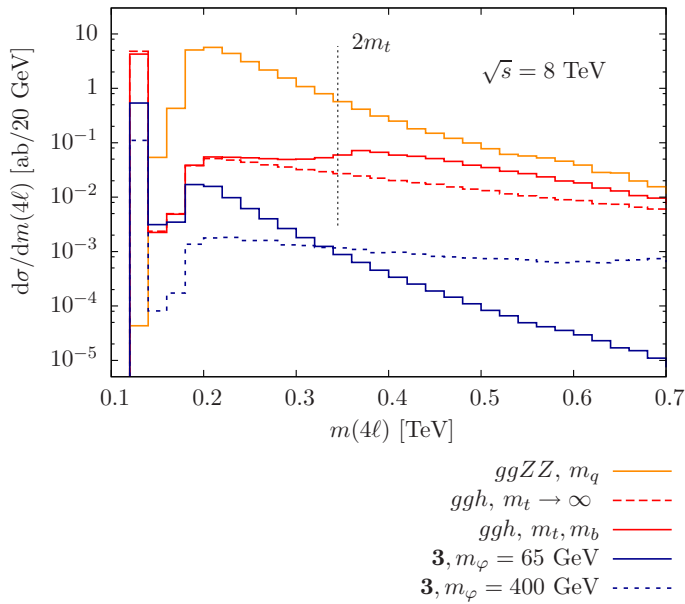}
    }
    \hspace{0.3cm}
    \parbox{0.4\textwidth}{ \vspace{3cm} \caption{Individual leading
        order contributions from Fig.~\ref{fig:feyzz} to the full
        hadronic cross section. For comparison we also include the
        effective theory distribution resulting from a $ggh$ effective
        vertex in the $m_t\to \infty$ limit. Cuts are identical to
        Fig.~\ref{fig:zzsm}. The coloured scalars are for representative
        values of $\lambda$ and $\Gamma_h$ to illustrate their
        behaviour. For additional details see text.}
      \label{fig:zz2}
      }
  \end{center}
\end{figure*}
%%%%%%%%%%%%%%%%%%%%%%%%%%%%%%%%%%%%%%%%%%%%%

Since the Higgs width is anticipated to be a small parameter compared
to the Higgs mass $\Gamma_h/m_h \sim 10^{-4}$, we can expand the Higgs
Breit-Wigner propagator ${\cal{D}}(s) = i/(s-m_h^2 + i\Gamma_hm_h)$
away from the peak region $s\gg m_h^2$
\begin{equation}
  \label{eq:prop}
  |{\cal{D}}|^2 = {1\over s^2} \left( 1 +  {m_h^4\over s^4}
    {\Gamma_h^2\over m_h^2} \right) +{\cal{O}}\left({\Gamma^4\over s^4}\right)\,
\end{equation}
which shows that the Higgs width parameter
rapidly decouples from the scattering process for Higgs off-shell
production. Therefore, the contribution from the triangle diagrams in
Fig.~\ref{fig:feyzz} (neglecting interference for the moment)
scales as
\begin{equation}
  \label{eq:prop2}
  {\text{d}}\overline \sigma_h \sim {g^2_{ggh}(\sqrt{s})\,
    g^2_{hZZ}(\sqrt{s})\over s}~\text{d}{\text{LIPS$\times$pdfs}}.
\end{equation}
Now, if there is a direct correspondence between $g_i(m_h)$ and
$g_i(\sqrt{s})$, measuring the signal strength $\mu$ in the off-shell
and on-shell regions simultaneously allows us to set a limit on the width
of the Higgs boson $\Gamma_h$.  More explicitly, for
$\Gamma_h>\Gamma_h^{\text{SM}}$, we need to have $g^2_{ggh} g^2_{hZZ}
> (g^2_{ggh} g^2_{hZZ})^{\text{SM}}$ to keep $\mu=\mu^{\text{SM}}$
fixed, which in turn implies $\overline\sigma_h > \overline\sigma_h^{\text{SM}}$.
Fig.~\ref{fig:zzsm} validates this line of thought and qualitatively
reflects the CMS analysis. 

\bigskip

But how general is this approach, or put differently, how solid is a
limit on $\Gamma_h$ obtained this way once we include unknown new
physics effects? And letting aside the interpretation in terms of a
constraint on the Higgs width, what are the more general ramifications
of a measurement of the gluon-fusion $ZZ$ and $WW$ cross section away
from the Higgs mass peak?

It is the purpose of this letter to address these questions from a new
physics perspective with a particular emphasis on probability
conservation.
First we interpret the outlined Higgs width measurement from a
unitarity perspective, which paves the way to the formulation of a
simple and transparent BSM counterexample. We analyse the interplay of
new resolved physics contributions to $gg \to VV^\ast$ to both Higgs
and continuum $ZZ, WW$ production in light of electroweak precision
constraints and finally point out that, enforcing $\mu\simeq
\mu^{\text{SM}}$ the off-shell measurement provides additional
statistical pull to constrain the Higgs' \cp~nature in the presence of
higher dimensional operators (unresolved new physics). We also discuss
off-shell measurements in WBF in Sec.~\ref{sec:wbf}.

As we will see, in order to gain qualitative control of new physics
effects in the Higgs off-shell region we cannot rely on effective
theory calculations for the SM spectrum. We consequently keep all
quarks dynamical and include finite mass effects of the bottom and top
quarks. Our work therefore extends beyond the assumptions of
Ref.~\cite{fermilab1} which has discussed the impact of new operators
to high invariant mass measurements in detail recently. We only focus
on modified $ggh$ and $hZZ/hWW$ interactions and neglect QED contributions
throughout; they are negligible for high invariant masses when both
$Z$s are fully reconstructed, but can be sensitive to the presence of
new physics when studied on the Higgs peak via $h\to
Z\gamma^\ast,\gamma^\ast\gamma^\ast$~\cite{fermilab2}. We will mainly
focus our discussion on $\sqrt{s}=8$~TeV; our results
straightforwardly generalize to run II.

Computations have been performed and cross checked with a combination
of {\sc{FeynArts/FormCalc/LoopTools}}~\cite{feyntools},
{\sc{Helas}}~\cite{Murayama:1992gi},
{\sc{MadGraph/MadEvent}}~\cite{Alwall:2011uj}, and
{\sc{Vbfnlo}}~\cite{vbfnlo}.  We have checked our results
against~\cite{ciaran} and find very good agreement.

\section{Higgs Width Measurements from $gg\to VV$: A Unitarity
  Perspective}
\label{sec:unit}

In Fig.~\ref{fig:zz2} we show the individual contributions of $pp\to
ZZ^\ast \to e^+ e^- \mu^+ \mu^-$ that result from the Feynman diagrams
of Fig.~\ref{fig:feyzz}. We also include a comparison of the full
Higgs contribution with the low energy effective theory~\cite{kniehl}
as implemented in {\sc{MadGraph/MadEvent}}~\cite{Alwall:2011uj}, which
shows large deviations when the absorptive parts of the top quark loop
are resolved (the corresponding Cutkosky cut~\cite{Cutkosky:1960sp} is
included in Fig.~\ref{fig:feyzz}). Obviously, a reliable analysis of
the high invariant mass region in correlation with the on-shell part
cannot be obtained by applying effective theory simplifications. The
CMS analysis~\cite{cmswidth} focuses on $m(4\ell)\geq 330$~GeV.

%%%%%%%%%%%%%%%%%%%%%%%%%%%%%%%%%%%%%%%%%%%%%
\begin{figure}[!b]
  \begin{center}
    \includegraphics[width=0.45\textwidth]{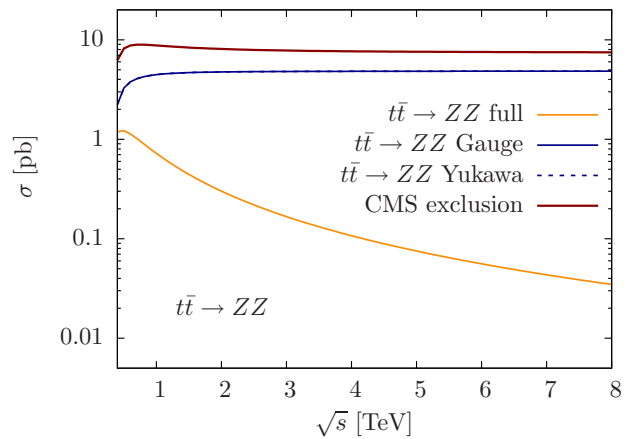}
    \caption{Unpolarized $t\bar t \to ZZ$ cross section as function of
      energy. We demonstrate unitarity cancellations between the gauge
      and Yukawa-type interactions (blue solid and dashed; the dashed
      line lies on top of the solid line), yielding a well-defined SM
      cross section (orange). We also show the parameter choice that
      corresponds to the CMS-like exclusion of $\Gamma_h \simeq 5
      \times \Gamma_{h}^{\mathrm{SM}}$ based on the strategy outlined
      in~\cite{hcms} and the introduction.}
      \label{fig:ttzz}
  \end{center}
\end{figure}
%%%%%%%%%%%%%%%%%%%%%%%%%%%%%%%%%%%%%%%%%%%%%

%%%%%%%%%%%%%%%%%%%%%%%%%%%%%%%%%%%%%%%%%%%%%
\begin{figure*}[!t]
  \begin{center}
    \includegraphics[width=0.45\textwidth]{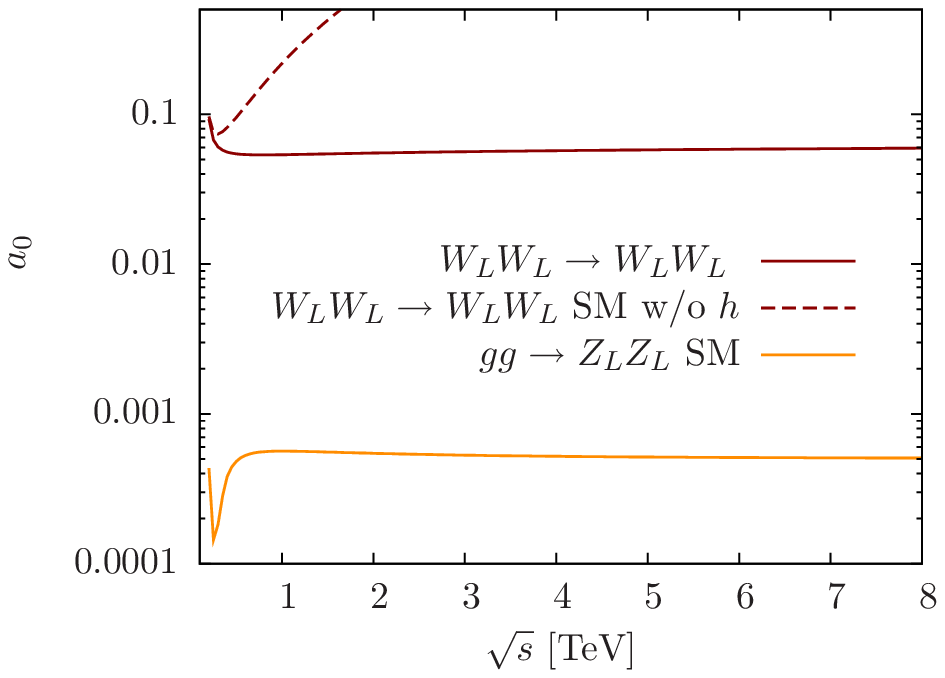}
    \hspace{0.6cm}
    \includegraphics[width=0.45\textwidth]{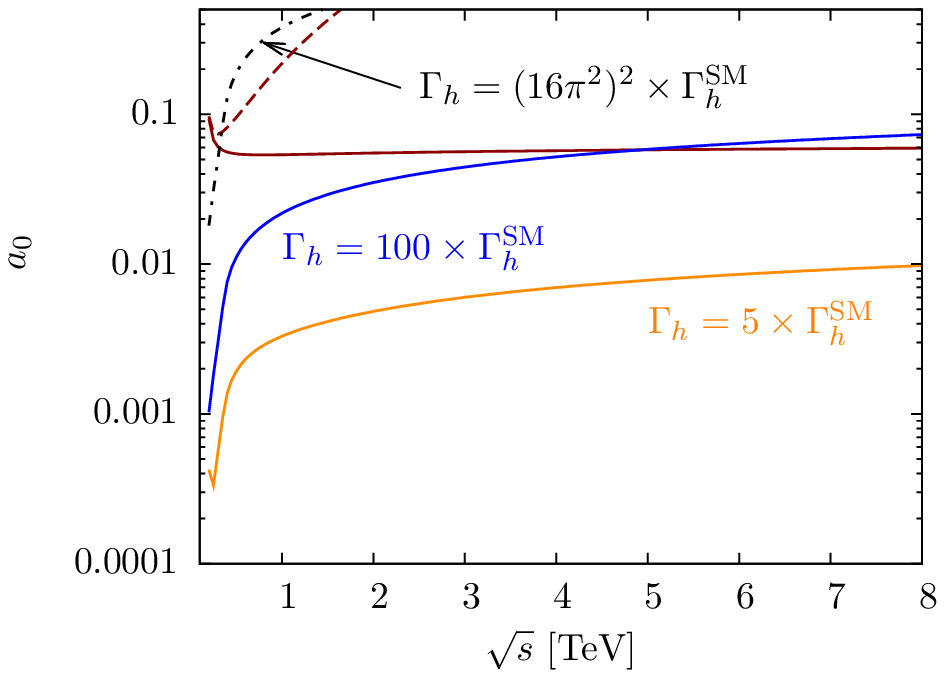}
    \caption{Zeroth partial wave projection for $gg\to Z_L Z_L$ for
      the SM and various values of the $g_{ggh}g_{ZZh}$-rescaling as a
      consequence of $\mu=1$ and $\Gamma_h>\Gamma_h^{\text{SM}}$. We
      also show the partial wave projection for longitudinal $WW$
      scattering in the SM with and without Higgs to put $gg\to ZZ$
      into context.}
      \label{fig:a0}
  \end{center}
\end{figure*}
%%%%%%%%%%%%%%%%%%%%%%%%%%%%%%%%%%%%%%%%%%%%%

It is known that the interference between the triangle and box
diagrams is destructive~\cite{Kauer:2013qba} above the $2m_t$
threshold. This large interference effect becomes transparent when
calculating the cross section for the process $q\bar q\to ZZ$ with
massive quarks in the initial state. It involves a highly
non-trivial cancellation between the gauge and Yukawa sector
interactions in $q \bar q \to Z_L Z_L$~\cite{Chanowitz:1978mv}, and is
part of the absorptive $gg \to ZZ$ amplitude according to the branch
cuts shown in Fig.~\ref{fig:feyzz}.  Even though both contributions
are gauge invariant under QCD transformations, they are related by
weak gauge invariance, and only their coherent sum with SM-like
couplings is well-defined. 

In Fig.~\ref{fig:ttzz} this is demonstrated for the unpolarized $t\bar t
\to ZZ$ cross section: The $s$ channel Yukawa couplings of $\bar t_L
t_R h+{\text{h.c.}}$ conspire via a coupling relation with the weak
gauge interactions $g_L\bar t_L\slashed{Z}t_L + g_R\bar
t_R\slashed{Z}t_R$ when $-t,-u \sim s$. A simple rescaling of one part
of the amplitude is tantamount to unitarity violation in the
Fermion-gauge interactions. This leaves a crucial question of the
limit obtained in~\cite{cmswidth}: Is the theory underlying the width constraint well defined?

The alert reader might object at this stage that such a question, in
fact, is also well-motivated for Higgs couplings measurements as
performed by ATLAS and CMS~\cite{Heinemeyer:2013tqa} when Higgs
couplings are varied independently throughout the SM Lagrangian. This
is certainly true if one would like to understand deviations from an
electroweak precision point of view. However, the situation for Higgs
$\sigma \times \BR $ phenomenology is fundamentally different. The
relevant scale at which couplings are evaluated is the Higgs mass and
$\sigma \times \BR $ phenomenology is manifestly free of UV problems
to leading order in the electroweak perturbative series
expansion.\footnote{This will dramatically change when the
  measurements of differential weak boson fusion distributions will be
  scrutinized at high precision~\cite{Ciccolini:2007ec}.} This needs to
be contrasted with an off-shell measurement that integrates over an
invariant-mass region $2m_t\lesssim m(4\ell) \leq
1.6~\text{TeV}$~\cite{cmswidth}.

To address this question quantitatively, we show the zeroth partial
wave projection as a function of the partonic center of mass energy in
Fig.~\ref{fig:a0}. Unitarity is violated when
$a_0>0.5$~\cite{Jacob:1959at}, and to contextualize our $gg\to ZZ$
findings with the SM Higgs sector we also show curves for SM $W_LW_L$
scattering that violates unitarity at low scales if the Higgs
contribution is neglected.

Indeed, the $gg\to Z_L Z_L$ scattering is sensitive to the coupling
rescaling as can be seen from Fig.~\ref{fig:a0}, however the partial
wave does not get close to $0.5$. The amplitude is sufficiently
diluted by loop factors $16\pi^2\sim 160$. Once this factor is
resolved the unitarity bound becomes relevant. This, however,
corresponds to a regime where the narrow width approximation is
violated entirely.

Although the limit in this channels is not afflicted with probability
non-conservation, it should be clear that the invoked rescaling leads
to an ill-defined electroweak sector as demonstrated in
Fig.~\ref{fig:ttzz}, the triangle and box contributions remain
intimately related. If high invariant mass measurements in the $gg \to ZZ$ channel 
yield a statistically significant increase over the SM, the
interpretation in terms of a modified Higgs width becomes
model-dependent. 

\section{Decorrelating on-shell and off-shell measurements in BSM
  theories}

\subsubsection{BSM contributions to Higgs production}
\label{sec:scalars}

The interplay in the absorptive parts of $gg\to ZZ$ linked by
unitarity in the high invariant mass regime and non-decoupling of top
loops tells us that the limit setting procedure outlined in the
introduction is based on a consistency argument for the electroweak
sector and is very specific to masses that are generated through
chirality-changing interactions. This paves the way to construct a
straightforward counterexample of the Higgs width measurement as
outlined above.

\begin{figure}[!t]
  \begin{center}
    \includegraphics[width=0.4\textwidth]{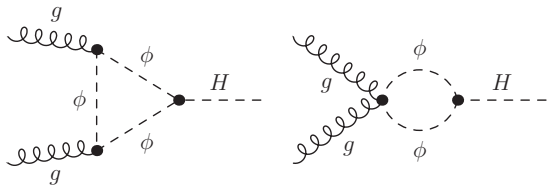}
    \caption{New Feynman diagram topologies to Higgs production via
      gluon fusion arising from Eq.~\eqref{eq:newlag}.}
      \label{fig:newfey}
  \end{center}
\end{figure}

Consider $\phi$, a scalar ${\bf{3}}$ under SU(3)$_C$, coupled to the
Higgs sector via portal interactions (see, {\it e.g.},
Ref.~\cite{Kumar:2012ww})
\begin{equation}
  \label{eq:newlag}
  {\cal{L}}_\phi = \left|D_\mu \phi \right |^2 - \tilde m_\phi^2 |\phi|^2 -
  \lambda |\phi|^2 |H|^2 + \dots \,.
\end{equation}
When the Higgs field obtains its vacuum expectation value $v$, the field $\phi$ induces a
contribution to single-Higgs production due to
the interaction $\lambda v |\phi|^2h$, as shown in
Fig.~\ref{fig:newfey}. The physical mass $m_\phi^2=\tilde m_\phi^2 +
\lambda v^2$ is essentially a free parameter $m_\phi^2>0$.

%%%%%%%%%%%%%%%%%%%%%%%%%%%%%%%%%%%%%%%%%%%%%
%%%%%%%%%%%%%%%%%%%%%%%%%%%%%%%%%%%%%%%%%%%%%

The new contribution gives an additional and potentially large
constructive or destructive contribution to $gg\to h$, depending on
the sign and size of $\lambda$~\cite{Dobrescu:2011aa}. To enforce
SM-like signal strengths $\mu\simeq 1$ we need to
introduce a compensating contribution to the Higgs width (this could
be interpreted as a Higgs-portal dark matter realization) and we have
$\Gamma_h > \Gamma_h^{\text{SM}}$.

Due to the scalar and electroweak singlet nature of the new fields we
only change the triangle Higgs production contribution while leaving
the box $gg \to ZZ$ contributions unaltered. Note, for this particle
there is no unitarity relation between the boxes and triangles. We
show the individual contributions of the scalar color triplet in
Fig.~\ref{fig:zz2}, which allows us to compare their behaviour with the
SM contributions. The scalar loops can easily be suppressed by two
orders of magnitude, leaving absolute and interference contributions
to the total hadronic cross section small for energetic events. This
behaviour is qualitatively known from supersymmetric
scenarios~\cite{Bonciani:2007ex} but has also been discussed in
non-supersymmetric models \cite{Kumar:2012ww,Dobrescu:2011aa};
effectively we have achieved a decorrelation of $g_{ggh}(m_h)$ and
$g_{ggh}(m(ZZ)>m_h)$, and the measurement can no longer be interpreted
as a Higgs width constraint.

To qualitatively understand why the scalars are suppressed at large
invariant masses, let us consider the ratio of the off-shell $gg\to h$
subamplitudes for scalars and tops (assuming
$m_\phi=m_t=y_tv/\sqrt{2}, \lambda= y_t$ for simplicity):
\begin{equation}
  \label{eq:ampr}
y_t\, {{\cal{M}}_\phi\over {\cal{M}}_t} = {1+2m_t^2 C_0(s,m_t) \over
  (s-4m_t^2) C_0(s,m_t)  - 2}\,,
\end{equation} 
where $C_0(s,m_t^2)$ denotes the characteristic scalar three-point
function following the Passarino-Veltman
reduction~\cite{Denner:1991kt}. The $\phi$-induced amplitude is
suppressed $\sim s^{-1}$, leading to a dominant behaviour of the top
loops at large momenta. This means that, even though we have a
modified Higgs phenomenology at around $m_h\simeq 125$ GeV it is exactly
the decoupling of the Higgs width according to Eq.~\eqref{eq:prop}
which renders the high invariant mass measurement insensitive to
modifications of $\Gamma_h$.

There is an interesting possibility when we consider larger $\phi$ masses and
larger couplings $\lambda$. For invariant masses $s^2\geq 4m_\phi^2$
we can have a sizeable constructive interference of the $\phi$ diagrams
with the top loops and as a result the cross section for large $m(4\ell)$
rises again and we recover the qualitative behaviour of
Ref.~\cite{melnikov}. For these parameter choices, however, we find
that the excess is smaller than expected for rescalings of
$g_{ggh}g_{ZZh}$ to keep $\mu\simeq 1$, Tab.~\ref{tab:xsec}. Similar
effects show up for light spectra $ m_\phi \lesssim 2m_t$, where this
interference in destructive and the high invariant mass search region
has a slightly smaller cross section although
$\Gamma_h/\Gamma_h^{\text{SM}}\gg 1$, outside the current CMS
exclusion.

%%%%%%%%%%%%%%%%%%%%%%%%%%%%%%%%%%%%%%%%%%%%%
\begin{table}
  \begin{tabular}{c | c | c | c }
    \hline
    $m_\phi$ & $\mu$ ($h$ peak) & $\Gamma_h/\Gamma_h^{\text{SM}}$ & $\overline \sigma/ \overline
    \sigma^{\text{SM}}$ [$m(4\ell)\geq 330$~GeV]\footnote{We impose the cut set used by CMS~\cite{cmswidth}
      without the {\sc{Mela}} cut~\cite{Chatrchyan:2012ufa}.}\\
    \hline
    70~GeV  & $\simeq 1.0$ &  $\simeq 5$  & $-2\%$\\
    170~GeV  & $\simeq 1.0$ &  $\simeq 4.7$  & $+80\%$\\
    170~GeV  & $\simeq 1.0$ &  $\simeq 1.7$  & $+6\%$\\
    \hline
  \end{tabular}
  \caption{Results for a single triplet scalar~\gl{eq:newlag} giving
    the correlation between $\mu$, $\Gamma_h/\Gamma_h^{\text{SM}}$ and
    high invariant mass cross section $\overline \sigma$ for the CMS
    selection cuts.}
  \label{tab:xsec}
\end{table}

In total, it is well possible to achieve $\Gamma_{h}\gg
\Gamma_{h}^{\text{SM}}$ without modifying the high invariant mass
regime of $pp\to 4\ell$ and without running into unitarity issues as
mentioned above. If such a contribution can be present the Higgs width
is an essentially unconstrained parameter, at least for a measurement
as outlined in~\cite{melnikov,cmswidth}.

%%%%%%%%%%%%%%%%%%%%%%%%%%%%%%%%%%%%%%%%%%%%%
\begin{figure*}[!t]
  \begin{center}
    \includegraphics[width=0.75\textwidth]{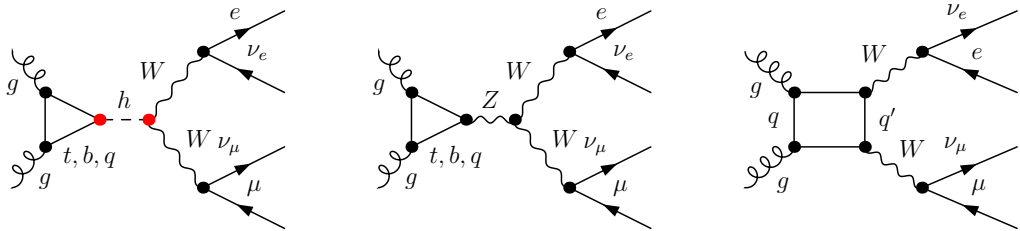}
    \caption{Representative Feynman diagram topologies contributing to
      $gg\to WW$ with leptonic $W$ boson decays. QED contributions are
      identical to zero due to Furry's theorem~\cite{Furry:1937zz},
      the $Z$ boson only probes an axial vector component.}
    \label{fig:feyww}
  \end{center}
\end{figure*}
%%%%%%%%%%%%%%%%%%%%%%%%%%%%%%%%%%%%%%%%%%%%%

Even though Eq.~\gl{eq:newlag} is a toy model to demonstrate the
limitations of total Higgs width measurements in the $gg\to 4\ell$
channel, color triplets of this form appear in any supersymmetric BSM
scenario and our argument has a broad validity, see {\it
  e.g.}~\cite{Bonciani:2007ex,Muhlleitner:2006wx} for a discussion of
squark contributions to Higgs production from gluon fusion in the
MSSM. If the extra scalars are charged under flavour, {\it e.g.} they
are top partners, exclusion will remain difficult \cite{excl1} for the
SUSY chimney regions (note, there are two chimney regions where one
can hide 170 GeV and 70 GeV scalars). Despite being color charged,
they could exist as stable particles on collider lifetimes when SUSY
is relaxed~\cite{excl2}. Quite naturally, details quickly become
highly model-dependent. By fixing $m_\phi$ we can map
$\Gamma_h=4.2\times \Gamma_h^{\text{SM}}$ onto $\lambda$ and obtain
$\overline \sigma /\overline \sigma^{\text{SM}}$ in
Tab.~\ref{tab:xsec}. When the $m_\phi$ masses become heavy we start
approaching an effective theory limit, which quickly decouples unless
we allow non-perturbative couplings as $m_\phi$ is not generated via
the Higgs mechanism. This is also visible in Fig.~\ref{fig:zz2} and we
recover the qualitative behaviour of Ref.~\cite{melnikov} also in this
model as alluded to above.  It is important to note, however, that the
interpretation of the measurement is still far more complicated. If we
imagine to become sensitive to the SM tail distribution within a small
error whilst observing a SM-like Higgs peak phenomenology, the mass
constraints to decorrelate the on- and off-shell region are weakened
and heavier coloured bosons in this channel and scenario become
essentially unconstrained.

Even though we have limited our discussion to $ZZ\to 4\ell$, the
findings of this sections straightforwardly generalize to $ZZ\to 2\ell
2\nu$ and $WW$.

\subsubsection{BSM contributions to continuum $ZZ$ and $WW$ production}

Our previous example shows that new contributions to $gg\to h$ can
significantly loosen the bounds on the Higgs width interpretation. In
a similar fashion we can imagine a situation where $\Gamma_h\neq
\Gamma_h^{\text{SM}}$ and the correlation of Eq.~\gl{eq:onshell} and
\gl{eq:prop2} is changed by new contributions to continuum $ZZ$
production. Such effects are expected in composite Higgs
scenarios with vector-like quarks~\cite{Lodone:2008yy} and typically
have non-trivial and non-diagonal electroweak interactions in the
extended flavour sector. In realistic models~\cite{Contino:2006qr}
such sectors can be quite large, and the phenomenology becomes
non-transparent especially when the different mass scales are resolved
and effective field theory simplifications cannot be applied. 

To keep our discussion as transparent as possible we will focus on a
minimal, anomaly-free toy model of vector-like quarks:
\begin{alignat}{5} 
  -{\cal L}\,\, &\supset m_Q \overline{Q}'_L {Q}''_R + m_d
  \bar{d}''_L {d}'_R
  + m_u \bar{u}_L'' {u}_R' + \text{h.c.} \notag	 \\
  \label{eq:vector}
  & +y_d' (\overline{Q}'_L H ) {d}_R' + y_u' (\overline{Q}'_L i\sigma^2
  H^\dagger) {u}_R'  + \text{h.c.} \phantom{\,,} \\
  & + y_d'' (\overline{Q}''_R H ) {d}_L'' + y_u'' (\overline{Q}''_R i\sigma^2
  H^\dagger) {u}_L'' + \text{h.c.} \,,\notag
\end{alignat}
where $Q_L',Q_R''=({\bf{3}},{\bf{2}},-{1\over 2})$,
$d''_L,d'_R=({\bf{3}},{\bf{1}},-1)$,
$u''_L,u'_R=({\bf{3}},{\bf{1}},0)$ under
SU(3)$_C$$\times$SU(2)$_L$$\times$U(1)$_Y$, {\it i.e.} we choose
lepton hypercharge quantum numbers for simplicity. Depending on the
relative size of the Yukawa coupling $y_i$ we can dial between the
modifications in the triangle and box contributions, while the box
contributions become sensitive to flavour changing charged and neutral
current interactions that follow from diagonalizing
Eq.~\gl{eq:vector}. In this sense, Eq.~\gl{eq:vector} reflects the
qualitative features of more realistic models and allows us to correlate
the off-peak cross section with oblique parameter constraints.

Depending on the size of (non-)diagonal couplings we can, in
principle, induce a new scale at large invariant $ZZ$ or $WW$ masses,
see Figs.~\ref{fig:feyzz} and~\ref{fig:feyww}. In case the masses of
these extra fermions are dominated by vectorial mass terms, oblique
electroweak constraints~\cite{Peskin:1991sw} are largely insensitive
to their presence. For $y_i\equiv 0$ we generically find $|S|\sim
10^{-3}$, $T=0$ and $|U|\sim 10^{-2}$ over a broad range of parameter
choices by explicit calculations. For pure vector-like terms we find
for the CMS search region described in Ref.~\cite{cmswidth}
(neglecting again {\sc{Mela}} cuts)
\begin{equation}
  \label{eq:zzbsm}
\Delta_{ZZ}^v  {\overline \sigma^{\text{BSM}} - \overline\sigma^{\text{SM}} \over
    \overline\sigma^{\text{SM}}} \bigg|_{ZZ} \simeq -4.8\%
\end{equation}
even for light physical fermion mass scales
\begin{equation}
  \label{eq:masspoint}
  (m_{d,1},m_{d,2},m_{u,1},m_{u,2}) = (300,200,400,300)~\text{GeV}\,,
\end{equation}
which are already under pressure from direct search exclusion limits.
The fermions quickly decouple for larger masses, owing to the dimension
eight structure of the resulting effective theory. We can also define a
high invariant $WW\to \ell\ell' 2\nu$ mass region that is
characterized by a large value of the transverse mass
\begin{equation}
  m_T^2= (p_T+\slashed{p}_T)^2 - ({\vec{p}}_T + \slashed{\vec{p}}_T)^2\,,
\end{equation}
and we choose $m_T\geq 300$~GeV in the following in addition to
$p_T(\ell)\geq 10$~GeV, $|y_\ell|\leq 2.5$, $\Delta R(\ell\ell')\geq
0.2$ and $\slashed{p}_T\geq 20$~GeV. As can be seen in
Fig.~\ref{fig:feyww}, $WW$ has a qualitatively different sensitivity
to this particular model class. We find in this region
\begin{equation}
  \label{eq:wwbsm}
\Delta_{WW}^v=  {\overline \sigma^{\text{BSM}} -\overline\sigma^{\text{SM}} \over
    \overline\sigma^{\text{SM}}} \bigg|_{WW} \simeq -3.8\% \,.
\end{equation}
for our example mass point Eq.~\eqref{eq:masspoint}.

%%%%%%%%%%%%%%%%%%%%%%%%%%%%%%%%%%%%%%%%%%%%%
\begin{figure*}[!t]
  \begin{center}
    \includegraphics[width=0.45\textwidth]{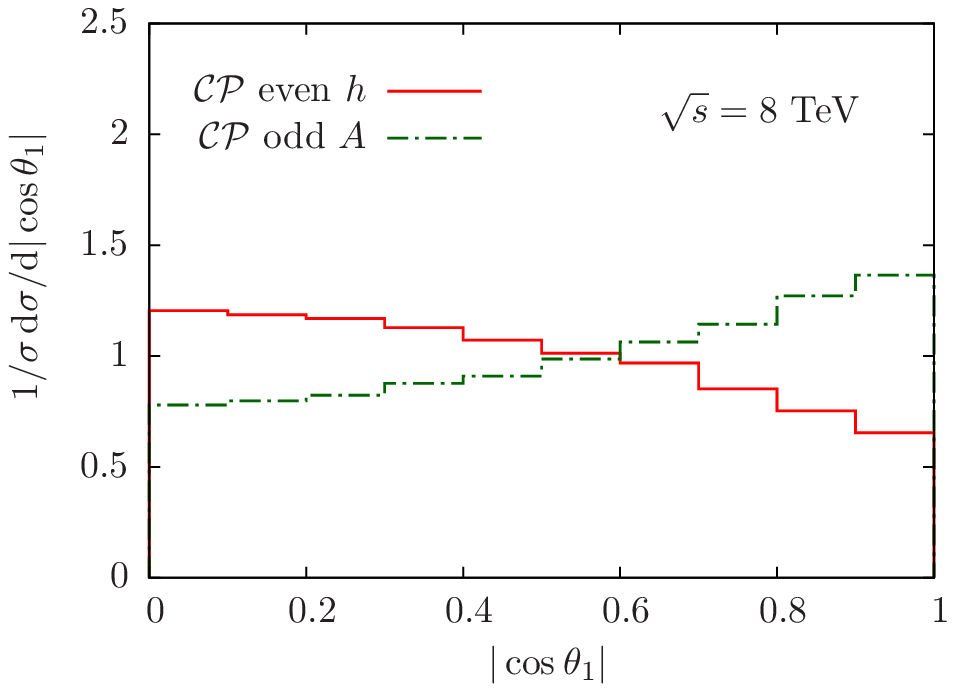}
    \hspace{0.6cm}
    \includegraphics[width=0.45\textwidth]{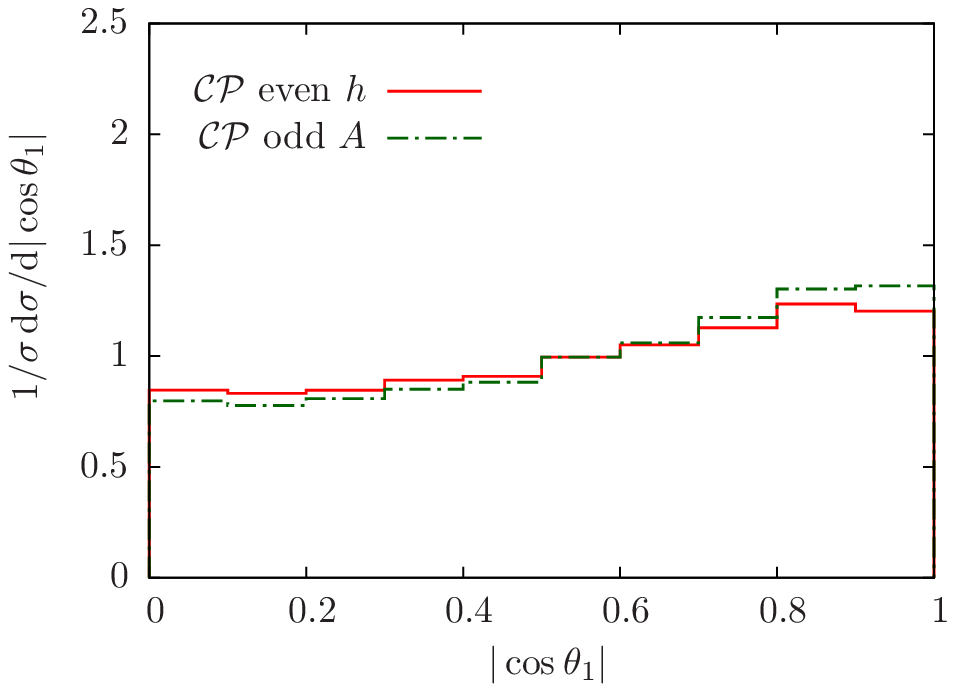}\\[0.3cm]
    \includegraphics[width=0.45\textwidth]{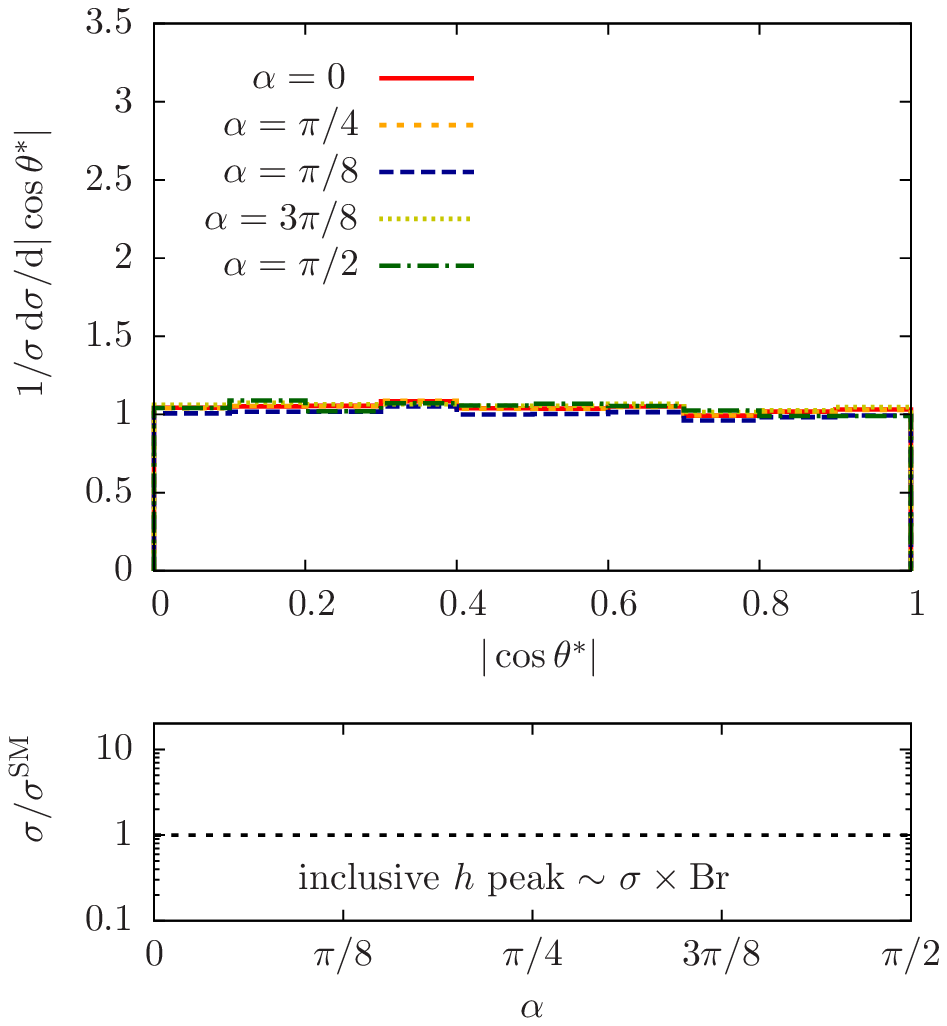}
    \hspace{0.6cm}
    \includegraphics[width=0.45\textwidth]{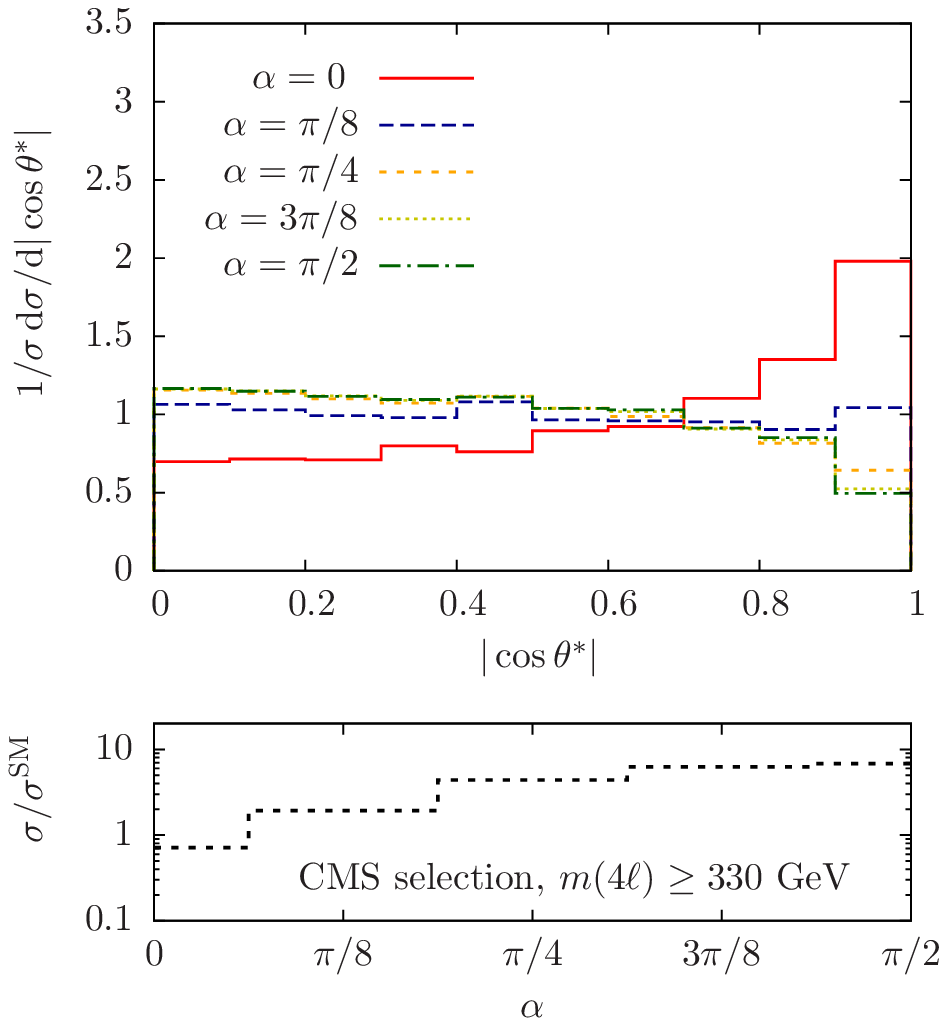}
    \caption{$\cos\theta_1$ (top panels) and $\cos\theta^\ast$
      distributions for the inclusive peak region and in the search
      region defined by CMS in Ref.~\cite{cmswidth}. $\alpha$ denotes
      the admixture of \cp~even vs. \cp~odd, signal strengths $\mu\sim
      1$ in the peak region are enforced via $\Gamma_h\neq
      \Gamma_h^{\text{SM}}$.}
      \label{fig:cstar}
  \end{center}
\end{figure*}
%%%%%%%%%%%%%%%%%%%%%%%%%%%%%%%%%%%%%%%%%%%%%

To understand how far oblique correction constraints limit the size of
novel electroweak degrees of freedom, we can use the above toy model
and take the mass spectrum of Eq.~\gl{eq:masspoint} as a
baseline. Then we change the ``chiral'' components by increasing
$y_i\geq 0$ until we reach the boundary of the $S,T$ exclusion
ellipse~\cite{fits}. For the resulting parameter point we obtain cross
sections analogous to Eqs.~\gl{eq:zzbsm} and~\gl{eq:wwbsm}
\begin{equation}
  \Delta^{v+h}_{ZZ} \simeq -4.3\%\,, \quad
  \Delta^{v+h}_{WW}\simeq -3.7\% \,.
\end{equation}

This example demonstrates that we can expect that large BSM
contributions to the continuum $ZZ$ production are highly limited by
electroweak precision constraints, especially because they also link
to non-trivial gauge representations under SU(3)$_C$.  The $ZZ$ and
$WW$ production modes are directly correlated with $S,T,U$ and Higgs
vev-induced mass terms in Eq.~\eqref{eq:vector} quickly introduce a
tension with electroweak precision constraints. They are known to have a
significant impact on $H\to \gamma \gamma$~\cite{Joglekar:2012vc} and
can be resolved at precision experiments~\cite{nloew}.

\section{Higher Dimensional Operators: Improving \cp~Sensitivity}
\label{sec:bsmg}
As a final application of the off-shell measurements we discuss the
impact of higher dimensional operators~\cite{dim6early,Alloul:2013naa}
on such a measurement. A recent
and comprehensive analysis has been presented in
Ref.~\cite{fermilab1}. Here we limit ourselves to the investigation of
\cp~properties, keeping all $m_t,m_b$ dependencies, and show that differential information in the high
invariant mass regime can be used to add a statistically independent
measurement to the \cp~hypothesis test as already performed by ATLAS
and CMS~\cite{cpexi}.  

For this purpose, we focus on interactions
\begin{equation}
  {\cal{L}} \supset \sum_{V=Z,W^+} c_{e,V}g_Vm_V V_\mu^\dagger V^\mu h
  + {c_{o,V}\over m_V^2}
  \tilde V^{\mu\nu} V_{\mu\nu} A
\end{equation}
and define the physical Higgs boson as a linear combination of
\cp~even and odd states,
\begin{equation}
X=\cos \alpha\, h + \sin \alpha\, A\,.
\end{equation}
We fix the signal strength for different angles $\alpha$ by changing
$\Gamma_h$ accordingly and focus in the following on the two angles
\begin{alignat}{2}
  \label{eq:cst1}
  \cos \theta_1 &= {{\vec{p}}(e^+) \cdot {{\vec{p}}}_{X} \over
    \sqrt{ {{\vec{p}}^2(e^+) }  {\vec{p}}_X^2} }\bigg|_{Z\to e^+e^-}\,, \\
  \label{eq:cstar}
  \cos \theta^\ast &= {{\vec{p}}(Z\to e^+e^-) \cdot {\vec{b}} \over
    \sqrt{ {{\vec{p}}^2(Z\to e^+e^-) }  {\vec{b}}^2} }\bigg|_X\,,
\end{alignat}
where $\dots|_R$ refers to the rest frame $R$ in which the angle is
defined. $p_\mu(X)=p_\mu(e^+)+p_\mu(e^-)+p_\mu(\mu^+)+p_\mu(\mu^-)$
coincides in the on-shell region with the Higgs boson's rest frame,
and $\vec{b}$ is an arbitrary three-vector along the positive beam
direction. As defined, $\cos\theta^\ast$ correlates the production
mechanism with the resonance's decay products by projecting onto the
beam-component of the $4$-lepton system.  While $\cos\theta^\ast$ is
known to be flat, $\cos\theta_1$ is sensitive to the ~\cp~properties
of the Higgs boson when produced in the on-shell region, see
Figs.~\ref{fig:cstar} and~Ref.~\cite{Gao:2010qx}. As can be seen, on
top of a cross section increase due to the higher dimensional operator
structure~\cite{fermilab1}, there is complementary information in the
spin/\cp~observables.\footnote{Not included in Fig.~\ref{fig:cstar} is
  the WBF contribution that can give rise to an additional $\sim 10 \%
  $ effect. We have checked the angular distributions with a modified
  version of {\sc{Vbfnlo}} and find no significant impact on the
  quoted results.}

\section{Off-Shell Measurements in Weak Boson Fusion}
\label{sec:wbf}
The potentially unknown loop contributions that can decorrelate the
on-shell and off-shell region in gluon fusion are not present in weak
boson fusion, assuming indeed a \cp~even SM-like Higgs boson. In these
channels, the method of Ref.~\cite{melnikov} becomes largely
model-independent except for a potential asymmetric deviation of the
$WWh$ and $ZZh$ couplings. This directly links to the $T$ parameter
and a deviation at tree level is expected to be small.

Furthermore, the weak boson fusion topology allows us to suppress
gluon fusion contributions using forward tagging jets in opposite
detector hemispheres with large invariant mass and rapidity gap
\cite{dieter}. By imposing an additional central jet veto
\cite{dieter2}, the gluon fusion events are almost entirely removed
from the sample \cite{DelDuca:2001eu} and the impact on a correlation
of the on- and off-shell regions will be unaffected by unknown physics
beyond the SM as a consequence.

\begin{figure}[!t]
  \begin{center}
    \includegraphics[width=0.45\textwidth]{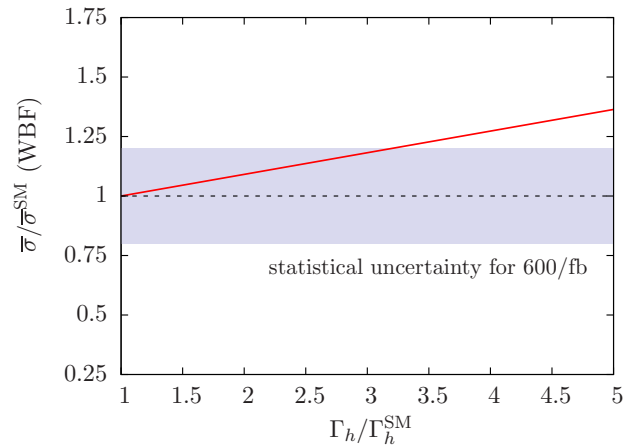}
    \caption{Weak boson fusion analysis of the off-shell measurement
      of Ref.~\cite{melnikov}. We apply hard weak boson fusion cuts to
      suppress a pollution from gluon fusion and include the
      statistical error based on a measurement with 600/fb. For
      details see text.}
    \label{fig:wq}
  \end{center}
\end{figure}

In Fig.~\ref{fig:wq}, we show the result of such an analysis at NLO
QCD \cite{jager,vbfnlo} (we choose a common rescaling of $g_{ZZh}$ and
$g_{WWh}$ to achieve $\mu\simeq 1$ in the on-peak region). The
selection cuts are identical to CMS' choice for the $Z$ reconstruction
and lepton selection. We lower the $4\ell$ mass cut to $m(4\ell) \geq
130$~GeV to increase the statistics as much as possible. In addition,
we employ typical WBF cuts \cite{dieter,dieter2,jager} as outlined
above
\begin{alignat}{3}
  p_{T}(j)>20~\text{GeV},~\Delta R(jj) \geq 0.6,~|y_j|<4.5,\notag\\
  \Delta y(jj)\geq 4.5,~y_{j_1}\times y_{j_2}<0,~m(jj)\geq 800~\text{GeV}\,,
\end{alignat}
and a jet veto 
\begin{equation}
  |y_j^{\text{veto}}| < 2.5,~p_T^{\text{veto}}(j)>50~\text{GeV},~\Delta y({j_{\text{veto}}j}) > 0.3\,.
\end{equation}
The leptons need to be well separated from the jets $\Delta R(\ell
j)\geq 0.6$ and need to fall inside the tagging jets' rapidity gap.
We furthermore reject events with $m(4\ell)>2$~TeV to avoid picking up
sensitivity from the region of phase space where the off-shell
modification probes the unitarity-violating regime.

Obviously, when performed in the WBF channel (our reasoning also
applies to the $WW$ channel), we observe a similar
behaviour~\cite{Kauer:2013qba}, however, at a much smaller cross
section $\overline\sigma \text{(WBF)} \simeq 0.04$~fb at 14 TeV
(already summed over light lepton flavours
$\ell=e,\mu$)~\cite{vbfnlo}. Nonetheless such a measurement can be
used to obtain a fairly model-independent measurement of the total
Higgs width following~\cite{melnikov} at large integrated luminosity,
especially when statistically independent information from multiple
WBF channels is combined.

%%%%%%%%%%%%%%%%%%%%%%%%%%%%%%%%%%%%%%%%%%%%%
%%%%%%%%%%%%%%%%%%%%%%%%%%%%%%%%%%%%%%%%%%%%%

%%%%%%%%%%%%%%%%%%%%%%%%%%%%%%%%%%%%%%%%%%%%%%%%%%

\section{Summary and Conclusions}
\label{sec:conc}

After the Higgs discovery with a mass of $m_h\simeq 125~\text{GeV}$
and TeV scale naturalness under siege, the total Higgs width is one of
the most sensitive parameters to light physics beyond the standard
model with a relation to the electroweak scale. A
\emph{model-independent} constraint on $\Gamma_h$ would have a huge
impact on BSM physics. Correlating on- and off-shell Higgs production
in $gg\to ZZ,WW$ and understanding cross section measurements in terms
of a total Higgs width limit, however, only applies to the
SM. Injecting the SM hypothesis into a global Higgs fit however will
always yield much tighter constraints~\cite{spans}.

The on- and off-peak continuum regions can be decorrelated in gluon
fusion by new degrees of freedom, which link to a modified Higgs
phenomenology; large BSM effects in continuum $gg\to ZZ,WW$ seem
unlikely given existing electroweak precision constraints.

If deviations at large invariant masses for $VV$ final states are
observed in the future, unitarity of the scattering amplitude dictates
that the interpretation of $\Gamma_h > \Gamma_h^{\text{SM}}$ will need
to involve model-specific assumptions, a fact that is unavoidable in
hadron collider physics. The cross section measurement can be used to
constrain momentum-dependent modifications of Higgs couplings that
underlie the modelling of spin/\cp ~testing and the general limit
setting procedure of higher dimensional operators.

Applying the strategy of Ref.~\cite{melnikov} to WBF allows us to
formulate a constraint that is largely free of the gluon fusion
shortcomings, however at considerably smaller cross sections.

A precise model-independent constraint on the total Higgs width
$\Gamma_h\simeq \Gamma_h^{\text{SM}}$, if not a measurement of
$\Gamma_h$ is challenging from a statistics and systematics point of
view and probably remains the remit of a future theoretically and
experimentally clean linear collider environment. At, {\it e.g.}, a
250 GeV $e^+e^-$ machine the combined investigation of associated and
WBF Higgs production, and exclusive final states $H\to b\bar b, WW$
allows us to formulate constraints on $\Gamma_h$ in the 10\%
range~\cite{Durig:2014lfa}.

\subsubsection*{Acknowledgments} 
CE is supported by the Institute for Particle Physics Phenomenology
Associateship programme.

%%%%%%%%%%%%%%%%%%%%%%%%%%%%%%%%%%%%%%%%%%%%%%%%%%
%%%%%%%%%%%%%%%%%%%%%%%%%%%%%%%%%%%%%%%%%%%%%%%%%%
%%%%%%%%%%%%%%%%%%%%%%%%%%%%%%%%%%%%%%%%%%%%%%%%%%

\end{document}